# Biomimetic temperature sensing layer for artificial skins

Raffaele Di Giacomo[1], Luca Bonanomi[1], Vincenzo Costanza[1], Bruno Maresca[2] and Chiara Daraio[1,3]

[1]Department of Mechanical and Process Engineering (D-MAVT), Swiss Federal Institute of Technology (ETH), Zurich, Switzerland.
[2]Department of Pharmacy, Division of Biomedicine, University of Salerno, Fisciano, Italy.
[3]Division of Engineering and Applied Science, California Institute of Technology, Pasadena, CA, USA.

**Artificial membranes that are sensitive to temperature are needed in robotics to augment interactions with humans and the environment and in bioengineering to improve prosthetic limbs. Existing flexible sensors achieved sensitivities of <100 mK and large responsivity albeit within narrow (<5 K) temperature ranges. Other flexible devices, working in wider temperature ranges, exhibit orders of magnitude poorer responses. However, much more versatile and temperature sensitive membranes are present in animals such as pit vipers, whose pit membranes have the highest sensitivity and responsivity in nature and are used to locate warm-blooded preys at distance. Here, we show that pectin films mimic the sensing mechanism of pit membranes and parallel their record performances. These films map temperature on surfaces with a sensitivity of at least 10 mK in a wide temperature range (45 K), have very high responsivity, and detect warm bodies at distance. The produced material can be integrated as a layer in artificial skins platforms and boost their temperature sensitivity to reach the best biological performance.**

Introduction

Artificial skins (*1, 2*) are essential to augment robotics (*3*) and improve prosthetic limbs (*4*). Existing platforms are designed to emulate properties of the human skin by incorporating sensitive functions (*4–8*) that respond to different external stimuli, e.g., to variations of temperature (*4, 8–10*). Available artificial skins that sense temperature variations use either passive, flexible resistors (*8–11*) or active electronic devices (*4, 12*). Their functionality is limited by the choice of temperature sensitive materials incorporated in the electronics (*2*). For example p-n junctions have small responsivity, require a complex architecture and demand for a non-trivial fabrication procedures (*4, 12*). Flexible sensors made of monolayer-capped nanoparticles are at the same time as sensitive to temperature as they are to pressure and humidity making impossible to deconvolve the three variables in practical applications (*8*). Composites based on a polymer matrix and electrically conductive fillers operate in a too narrow temperature range and have 2 orders of magnitude uncertainty on the current value corresponding to the same temperature (*9, 13*). Significant advances on artificial skins require the use of new flexible materials with higher temperature sensitivity, responsivity, range of operation and stability.

Recently, it has been shown that materials composed of plant cells and carbon nanotubes have very high responsivity over large temperature ranges (*14*). However, these materials are not suitable for artificial skins, since they have mechanical properties similar to wood, are not flexible and require cumbersome fabrication approaches. In this work, we focus on the active molecule responsible for the large temperature responsivity in plant cells (*pectin*) (*14*) and engineer films suitable for flexible electronic devices.

Pectin, a component of all higher plant cell walls, is made of structurally and functionally very complex, acid-rich polysaccharides (*15*). Pectin plays several roles in plants, for example, it is an essential structural component of cell walls that binds ions and enzymes (*16*). In high-ester pectins, at acidic pH, individual chains are linked together by hydrogen bonds and hydrophobic interactions. In contrast, in low-ester pectins, ionic bridges are formed, at near neutral pH, between $Ca^{2+}$ ions and the ionized carboxyl groups of the galacturonic acid, forming an "egg box" in which cations are stored (*17*). Since the crosslinkings between pectin molecules decrease exponentially with temperature (*18*), increasing the temperature of a $Ca^{2+}$-crosslinked pectin increases ionic conduction (*14*).

**Results and Discussion**

We produced pectin films and compared their temperature responsivity (i.e., the signal variation in a given temperature increment) with the best flexible temperature sensing films (*4*, *8*, *10*), in a biologically relevant 45 K temperature interval (Fig. 1A). The pectin films (Fig. 1B) signal variation is ca. 2 order of magnitude greater than the others (*4*, *8*, *10*). A higher responsivity has been reported only for a very narrow temperature range (<5 K) for a two-state device (*9*) (i.e., a temperature switch).

To find a closer match to the pectin film's responsivity, sensitivity and range of operation it must be compared directly to biological membranes. Human skin, for example, senses temperature with a sensitivity of 20 mK (*19*) through ion channels (*20*) that belong to the family of the TRP sensors and include the snake TRPA1 orthologous (*20*), which is the most sensitive temperature sensor in nature. Snakes are cold-blooded animals and their body temperature corresponds to that of the environment. Snakes' pit membranes (*21*) distinguish minute temperature variations. The extraordinary sensitivity of pit membranes is due to the presence of voltage-gated ion channels orthologues of the wasabi receptor (*21*) in humans. At night, thermal emission from a mammalian prey at a maximum distance of 1 m away causes a small, local temperature increase on the membranes (*21*). The small temperature increase leads to an increased opening of TRPA1 ion channels (*20–22*) and to an increased current carried by $Ca^{2+}$ ions (*21*) (Fig. 1C) through the cell membrane. Interestingly, the mechanism of detection (*23*) of the pit membrane is not photochemical since the incident thermal radiation is not converted directly into electrical current[1]. For this reason, the pit membrane response to temperature has been characterized measuring the current variation when placed in contact with a warm surface, *i.e.,* as a thermometer film rather than an optical receiver in the far infrared range (*21*). So far, no engineered material with similar thermal sensitivity or responsivity in a comparable range of temperatures has been reported. Here we show that pectin films mimic the mechanism of the TRP receptors by using a similar $Ca^{2+}$ current regulation (Fig. 1D) and achieve the same sensing performance of snakes' pit membranes (Fig. 1E).

We fabricated films (~200 μm thick) by casting a pectin solution in a mold (see Methods), thinner films can be produced by spin coating. The pectin was crosslinked in a $CaCl_2$ solution and dehydrated in vacuum to obtain a transparent film. After dehydration, the conductivity of the hydrogel is in the order of 0.1 mSm$^{-1}$. The current-voltage characteristic of a typical film is linear (Supplementary Fig. S1). To characterize the response of the material to temperature, samples' current was measured between 10 and 55 °C on a Peltier element. The temperature was monitored with an independent calibrated Pt100 sensor. The thermal responsivity achieved was comparable with that of rat and rattle snakes' pit membranes Fig 1E. The responsivity was also within the same order of magnitude of the plant cells-carbon nanotubes composites (*14*). However, the produced pectin films are transparent, flexible and conformable to any surface, thus ideal as sensitive layer in synthetic skins.

To prove their sensing mechanism, we made three control experiments measuring the temperature responsivity of (i) pure water, (ii) pectin films with pure water and no crosslinking ions, and (iii) a $CaCl_2$ solution. The temperature response in the three cases was much lower than that of the $Ca^{2+}$-crosslinked pectin (see Supplementary Fig. S2). This proved that the large responsivity of the crosslinked pectin films is due to interactions between pectin chains and the $Ca^{2+}$-ions as shown in Fig. 1D and reported in (*14*). To measure the stability of the $Ca^{2+}$ crosslinked pectin films, we cycled them in a 30 K interval Fig. 2A. The films are very stable over the 215 cycles tested (Fig. 2A), showing no significant change of responsivity (Fig. 2B) nor of absolute current values at each temperature (Fig. 2C). The activation energy for pectin films is 81.9 kJ/mol (Supplementary Fig. S3a,b and Supplementary Materials). A similar value of the pectin activation energy was reported in rheological measurements (*18*).

We tested the sensitivity of the pectin films, monitoring the local temperature of a sample (Fig. 2D) with a thermal camera, while an independent source-meter measured its electrical current. The film responds with fidelity to small changes of the environmental temperature (Fig.s 2D,E). The detailed data for a 2-sec time interval (Fig. 2E) reveal that the film senses temperature variations of at least 10 mK. To compare the performance of our material to that of the viper's pit membrane, we characterized the sensitivity of pectin films when facing small, warm bodies at a distance. A microwavable teddy bear was heated up to 37 °C. A thermal camera was used to determine its temperature and ensure it remained constant during the measurements. We placed the teddy bear 1 m from the membrane for ca. 20 sec and then removed it. We repeated this procedure also at distances of 0.6 m and 0.4 m. The results show that the membrane detects warm bodies, about the size of a rat or a small rabbit, at a distance of 1 m (Fig. 2F).

We also performed experiments with larger films (21 x 29.7 cm) connected to two carbon electrodes and operating at 20 V (see Methods, Supplementary Video 1 and discussion). These samples have the same sensitivity as measured in smaller films (Fig. 1E). To test the response of the films to bending, we monitored their current at constant temperature in different bending positions (Fig 3A and Supplementary Fig. S4). The current variations due to bending are negligible, compared to the variations induced by small temperature changes (Fig 3A). We also tested the response of a bent sample to temperature variations (Fig. 3B) and found no change in responsivity. The experiments in Fig 3B were performed on a copper bent substrate covered with and insulation layer.

To verify that pectin films can be integrated in a synthetic skin as a temperature sensitive material, we fabricated samples (52 mm × 52 mm) with multiple electrodes (8 or 16 contacts) deposited on the external frame (Supplementary Fig. S5A,B). These samples were made of pectin films with chromium/gold electrical contacts sandwiched between two insulating layers, to protect them from the effect of direct contact with external conductors and/or humidity/water (Fig. 3C). We monitored the signal between electrodes of each row and column, while increasing the temperature in selected areas of the skins. Based on the number of contacts on the outer frame, we divided the area of the sample in four (or sixteen) blocks, corresponding to the number of "pixels". Each pixel is addressed as the intersection between each row and column, according to the electrodes' position. This arrangement allowed us to reconstruct the temperature map on the material without cumbersome or pixel addressed electronics (*3*, *9*). Mapping of complex temperature profiles can be further enhanced by algorithmic analysis (*24*, *25*).

The measurements obtained in the 4-pixel sample were performed using the circuit shown in Supplementary Fig. S6 operating at 18 V and explained in the SM file. The position of a finger touching 4 different pixels of a skin for ~2 sec is clearly distinguishable from the electrical response of the materials (Fig. 3D). The voltage signals acquired are reported in Fig.s 3D, Supplementary S7 and Tab. S1. The noise in Supplementary Fig. S7 derives from the electronic readout circuit and not from the sensor as confirmed performing similar measurements with a pico-amperometer (Supplementary Fig. S8). The temperature variation on each pixel was c. a. 1 K, as shown in the thermal image in Supplementary Fig. S9. To exclude piezo-resistive effects, we performed the same measurements pressing the sample with a metal object at the same temperature of the pixel (Supplementary Fig. S10). To test the response of pectin skins to an increased sensing spatial density, we fabricated a 16-pixel device in the same skin area and evaluated its temperature mapping ability. We placed near the lower right corner of the skin an aluminum parallelepiped (12 mm x 12 mm x 3mm) at 26 °C (with an ambient temperature of 20 °C). As shown in Fig. 3E, we measured the signal on the skin for each of the 16 pixels, 0.8 sec after the aluminum square was laid in contact, (Supplementary Tab. S2). The thermal camera map (Supplementary Fig. S11 and pixelated in Fig. 3F) and the temperature map obtained with our skin show an excellent match.

The exquisite temperature sensitivity and mapping ability of pectin skins reveal opportunities in robotic sensing and haptics, where biomimetic sensors are important (*2*). For example, pectin skins could be embodied in robotic prosthetics, which are limited today by the need of improved sensory feedback (*26*). Feedback from prosthetics is essential to restore the complete functionality of a limb and is especially critical for achieving proper control of robotic devices, attaining much better results than with the single use of vision (*27*). Pectin skins can be used as a high-performance layer in flexible electronic devices, for example, when sandwiched in the architecture of an artificial skin or a prosthetic limb.

The record-high sensitivity of pectin skins make them suitable to record finely distributed temperature maps on surfaces. Their ease of fabrication and minimal requirements for electronic circuitry make them compatible with most existing flexible technologies. Some limitations arise from the need of an insulating layer against excessive humidity. The insertion of a polymeric insulation layer in synthetic skins is a common practice and can offer a direct solution to the problem. Another limitation is the need for accurate initial calibration. Improving the uniformity of the pectin layers is expected to reduce the current calibration complexity.

**Conclusions**

The present work demonstrates that a material composed exclusively of purified plant pectin and crosslinking ions, engineered into a film, has a performance equivalent to that of the snake's pit membrane and superior to other flexible materials. The pectin films are ultra-low cost and scalable, insensitive to pressure and bending and can be used to augment temperature sensing when integrated in synthetic skin platforms.

**Materials and Methods**

To produce the materials, we used commercially available citrus low-methoxylated pectin (LMP) with a degree of methylation of 34% and a content of galacturonic acid of 84% (Herbstreith&Fox©). Pectin powder (2% w/vol) was dissolved at 80 °C in deionized water and stirred at 1,400 rpm until a uniform solution was obtained. To jellify the films, a 32 mM $CaCl_2$ solution was prepared (corresponding to a stoichiometric ratio R = $[Ca^{2+}]/2[COO^-]$ = 1). The pectin solution was poured into a mold and the $CaCl_2$ solution then added. After gelation, the highly hydrated films were transferred to a vacuum chamber and dehydrated at 12 mbar overnight. Samples were then detached from the Petri dish using a razor blade. The large samples shown in Supplementary Video 1 were produced pouring the gel on a glass substrate (28 cm × 30 cm × 0.5 cm) as the lower insulating layer. The electrical contacts were made of carbon tape, and a clear insulating acetate sheet (A4 paper format) was layered on top. To produce the skins, the pectin solution was deposited directly on different substrates (PDMS, cellophane or $SiO_2$) with pre-deposited electrical contacts made by sputtering chromium/gold or using carbon tape. In the experiments described in Fig.s 1 and 2 a d.c. polarizing voltage of 20 V was applied to the samples and the current allowed to decrease for ca. 2 hours. After the initial discharge, the current remained stable for several hours (during which experiments where performed). The current was measured with a Keithley 2336B source meter. For the experiments in Fig. 3 the applied voltage was a square wave with an amplitude of 20 V and a frequency of 5 Hz. Sampling rate was 10 samples per second. Temperature on the film was actuated by a Peltier element QC-31-1.4-8.5M. Independent temperature measurements on the film were measured with a Pt100 platinum thermometer. We also performed measurements at different frequencies (see Fig. S12) up to 50 °C. We found no difference in the temperature response of the pectin films under a.c. or d.c. conditions (see Supplementary Materials).

**References.**

**Acknowledgements:**
The authors thank Ueli Marti (ETH Zürich) for the technical support and useful discussions. R.D.G., B.M., C.D., L.B., and V.C. are inventors on patent applications EP15161042.5, EP15195729.7, PCT/EP2016/056642 submitted by ETH Zurich that cover Gel based thermal sensors.

**Funding:**
This work was supported by the Swiss National Science Foundation, Grant #157162.

**Author contributions:**
R.D.G., L.B., V.C., B.M., and C.D. conceived the system and designed the research. R.D.G., L.B and V.C. designed and performed the experiments. R.D.G. and V.C. designed the readout circuit for the electrical measurements. All authors contributed to the analysis of the data and discussions. L.B. and V.C. prepared the figures. R.D.G, V.C., L.B. and C.D. designed the supplementary video. L.B. edited the video. R.D.G. and C.D. wrote the paper.

**Additional information:**
Supplementary Material is available in the online version of the paper.

**Competing financial interests:**
The authors declare no competing financial interests.

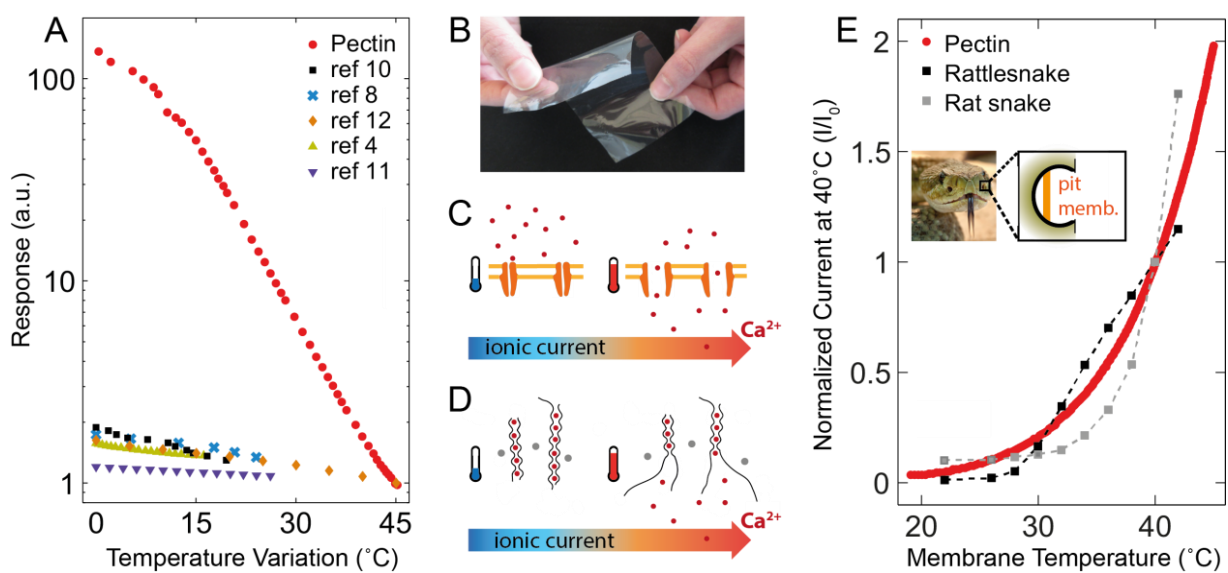

**Figure 1 | Comparison between artificial skins, snakes' pit membrane and pectin films. A,** Response of different artificial skins. Normalized signal variation as a function of relative temperature change. Red dots: pectin film resistance. Black Squares: resistance replotted from Park *et al. (10)*. Blue crosses: resistance replotted from Segev-Bar *et al.* (8). Orange diamonds: resistance replotted from Trung *et al. (12)*. Green triangles: voltage replotted from Kim *et al. (4)*. Violet triangles: resistance replotted from Webb *et al. (11)*. **B**, Digital image of a sample of the produced pectin films. **C,** Molecular mechanism governing pit membrane sensitivity. Dark orange: TRPA1 channels. Light orange: cell membrane. Red dots: $Ca^{2+}$ ions. **D,** Molecular mechanism governing crosslinked pectin films' sensitivity. Black lines: galacturonic acid. Red dots: $Ca^{2+}$ ions. Grey dots: water molecules. **E,** Comparison with pit membranes: Red dots: normalized current in a pectin film. The points with error bars and dotted lines are plotted from Gracheva *et al. (21)*. Dark grey dots: rattle snake. Light grey dots: rat snake. Inset: Digital image of a rattle snake and schematic of the pit organ and pit membrane.

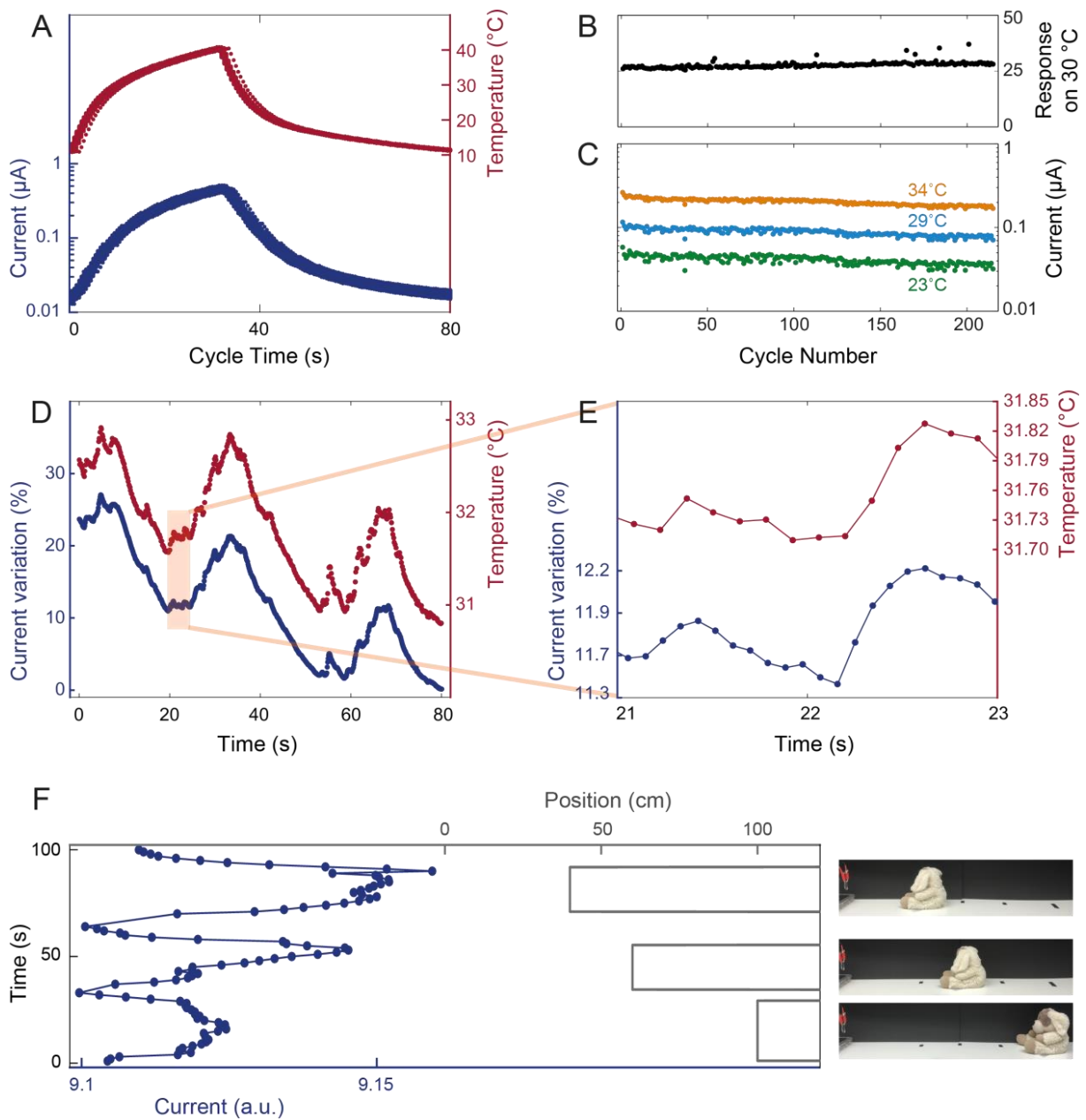

**Figure 2 | Characterization of the pectin films. A,** Red dots: forced temperature, 215 cycles superimposed. Blue dots: corresponding electrical current in a pectin film, 215 cycles superimposed. **B,** Responsivity: electrical current ratio between 40 and 10 °C during the 215 cycles displayed in (A). **C,** Electrical current in the pectin film at different temperatures during the 215 cycles displayed in (A) **D,** Electrical current value in the pectin film (blue dots, left axis) plotted as a function of time and compared to the sample's temperature measured by the thermal camera (red dots, right axis). The temperature oscillations are caused by variations of the ambient temperature during the measurements. **E,** Magnification of the data in the pink box in panel (D)**,** the dots: measurement points, lines are included as a guiding reference. **F,** Sensing heat from a warm object (37 °C) at a distance. Blue dots: electrical current in the pectin film, blue line is included as a guiding reference. Black line, position of the object with respect to the membrane positioned in 0.

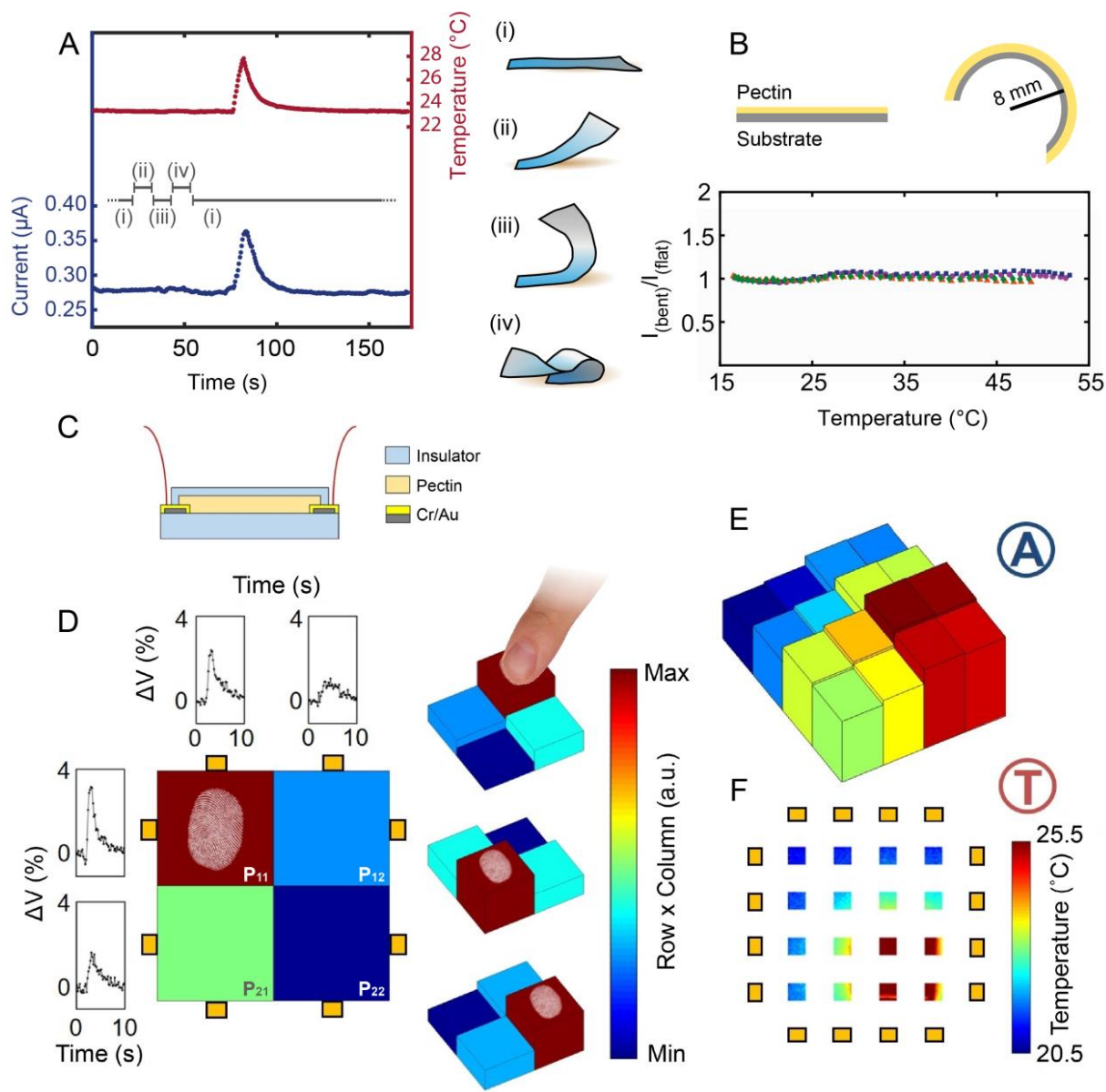

**Figure 3 | Characterization of the pectin films as materials for artificial skins. A,** Current and temperature as a function of time while bending. At t = 75 sec the temperature was increased and then decreased by ca. 4 K. On the right, cartoons of the bending positions tested, see Supplementary Fig. S4 for pictures. **B,** Current response when the sample is bent. Cartoon of the bending position of the film on a copper/insulator substrate. **C,** Schematic view of the pectin skins in cross section. **D,** Electrical response and temperature maps obtained with a 4-pixel skin, when a finger touched it in different positions (refer to the finger print location in each panel). The voltage-time panels show the signal readout for the corresponding rows and columns. The colors (and heights of the blocks) correspond to the product between the maximum signal variations (in %) detected in each row and column (see Supplementary Materials, Tab. S1), normalized to 1. Supplementary Figure S7 shows the percentage increase of the signal in time, for each of the 4-pixels when individually touched. **E,** Electrical response of a 16-pixel skin when a warm object is placed on its bottom right corner. **F,** Pixelated thermal camera image of the skin corresponding to (E).

# Supplementary Material

# Biomimetic temperature sensing layer for artificial skins

Raffaele Di Giacomo[1], Luca Bonanomi[1], Vincenzo Costanza[1], Bruno Maresca[2] and Chiara Daraio[1,3]

[1]Department of Mechanical and Process Engineering (D-MAVT), Swiss Federal Institute of Technology (ETH), Zurich, Switzerland.
[2]Department of Pharmacy, Division of Biomedicine, University of Salerno, Fisciano, Italy.
[3]Division of Engineering and Applied Science, California Institute of Technology, Pasadena, CA, USA.

## Supplementary Materials and Methods

### Polymeric insulation

We utilized polymeric insulation layers such as acetate (polyvinyl acetate) to protect the sensing layer from humidity and pH variations. This is a common practice in artificial skins. No chemical interaction between pectin and polyvinyl acetate or PDMS is expected due to their stable polymerized state. No change in responsivity or sensitivity was found with respect to pectin films without insulating layer when acetate or PDMS were used. Any other insulating material already in use for synthetic skins would serve for the scope.

### Measurements

The electrical measurements reported in Fig. 1, 2, 3A,B, S1, S2, S3, S8, S10 were performed in a two-point contact geometry using a source measurement unit (Keithley model 2635), also referred to as amperometer or pico-amperometer in the main text of the paper. The electrical measurements in Fig. S12 were acquired with a lock-in amplifier model SR830 Stanford research systems. For the electrical measurements reported in Fig. 3D,E and S7, we applied sequencially a square wave voltage having an amplitude of 18V to the electrical contacts in each row and column. We measured the signal output with the readout circuit (in Fig. S6), connected to a DAQ board (National Instruments® BNC-2110). The thermal camera used in the experiments was a FLIR® A655sc.

### Sensor's response and comparrison

"he metrological quantity of choice for the comparison in figure 1A is the response/responsivity defined as the amount of change in the output (readout signal) for a given change in the input (in this case temperature). The scale in the plot is the same for all the sensors. Each value on the plot can be calculated as (Output$_{T2}$)/(Output$_{T1}$) for each T2-T1 and with T1 fixed. The values were taken from the references cited, as reported in the legends and in the captions. Since the plot is in Log scale, any arbitrary scaling factor will result in a translation of the curves up or down along the y-axis, but the slope of the curves (response) will be preserved.

# Pectin

Pectin is composed of galacturonic acid for approximately 70% and all pectic polysaccharides contain galacturonic acid linked at the O-1 and the O-4 positions. Studies based on pectin biosynthesis have shown that pectin is synthesized in the Golgi apparatus and transported to the cell wall inside membrane vesicles (*1*). Once synthesized polymers move to the cell wall by the movement of Golgi vesicles, possibly along actin filaments through their myosin motors (*1*). The observed heterogeneity in the pectin structure is due to species-, cell type-, and developmental state-specific differences in enzyme composition (*2*). Essentially, all studies on pectins have been performed on economically important plants such as apple, citrus, sugar beet, tomato or on cultured cells as sycamore, carrot, spinach, and rose. Thus, very little is known of the 235,000 remaining known flowering plants and of pectins of green algae, liverworts, mosses, and ferns. In addition, characterization of pectin from cell wall mutants (*3*, *4*) and from plants growing in extreme environments may reveal unknown pectin structures with different properties. This variety suggests the large potential of pectin in future studies and synthesis of temperature sensitive layers. Pectin biosynthesis probably requires at least 67 transferases (glycosyl-, methyl-, and acetyl-transferase). It includes homogalacturonan, rhamnogalacturonan I, and the substituted galacturonans rhamnogalacturonan II (RG-II), and xylogalacturonan (XGA) (*5*). In low-ester pectins, the gelation that occurs within the egg-box is due to the presence of calcium ions can be quickly reversed by monovalent sodium and potassium cations (*6*). An initial dimerization step of two homo-galacturonic chains, by cooperative bridging through $Ca^{2+}$ ions determines the binding of the first calcium cation to two pectin chains facilitating their alignment with respect to each other, allowing an easier binding of an upcoming calcium ion (*7*).

# Pectin film

### Current-voltage characteristic

Figure S1 shows the current-voltage characteristic of a typical pectin film. The characteristic is linear and does not contain mayor bumps, steps or discontinuities. This is generally a sign of absence of major electrical defects in the bulk of the material or contacts' defects.

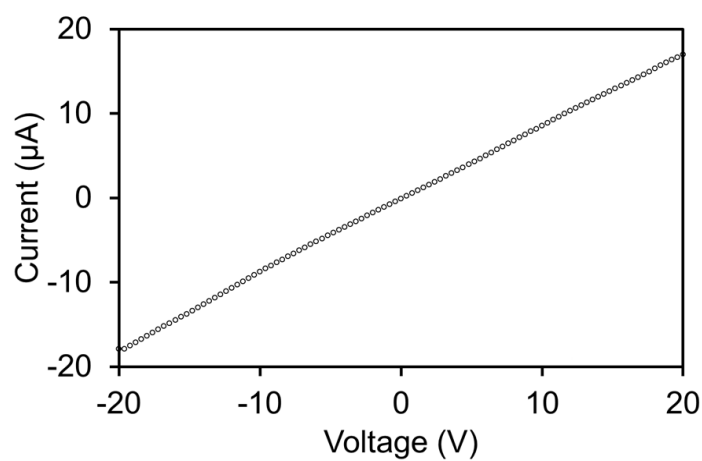

**Figure S1│Pectin film.** Current vs. voltage characteristic of a pectin film. Each circle represents a measurement point.

**Control experiments**

Figure S2 shows the conductivity as a function of temperature of three different samples: $Ca^{2+}$ crosslinked pectin films, pectin films with water but without crosslinking ions, and 32mM $CaCl_2$ solution.

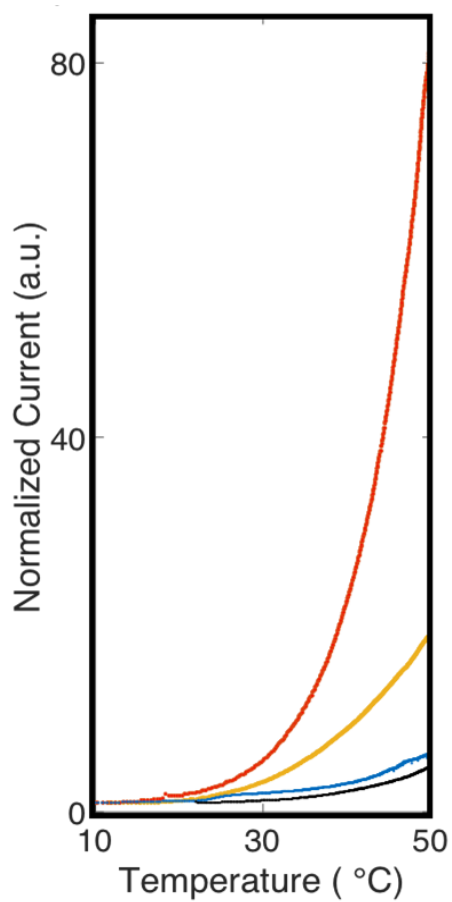

**Figure S2 Conductivity variation as a function of temperature, control experiments:** red dots $Ca^{2+}$ crosslinked pectin. Dark orange dots: pectin. Blue dots: pure water. Black dots: 32mM $CaCl_2$ solution.

**Current-temperature characteristics**

Figure S3a shows the temperature and current measurements as a function of time for a tipical pectin film. The experiment was performed by increasing the temperature of the film from 8 to 30 °C and then from 30 to 39 °C using a hotplate. Temperature was monitored using a thermal camera (FLIR A655sc) directly pointing to the film. The current across the device was monitored using a source meter (Keithley model 2635) and followed the temperature profile measured by the thermal camera. The temperature-current relation is exponential.

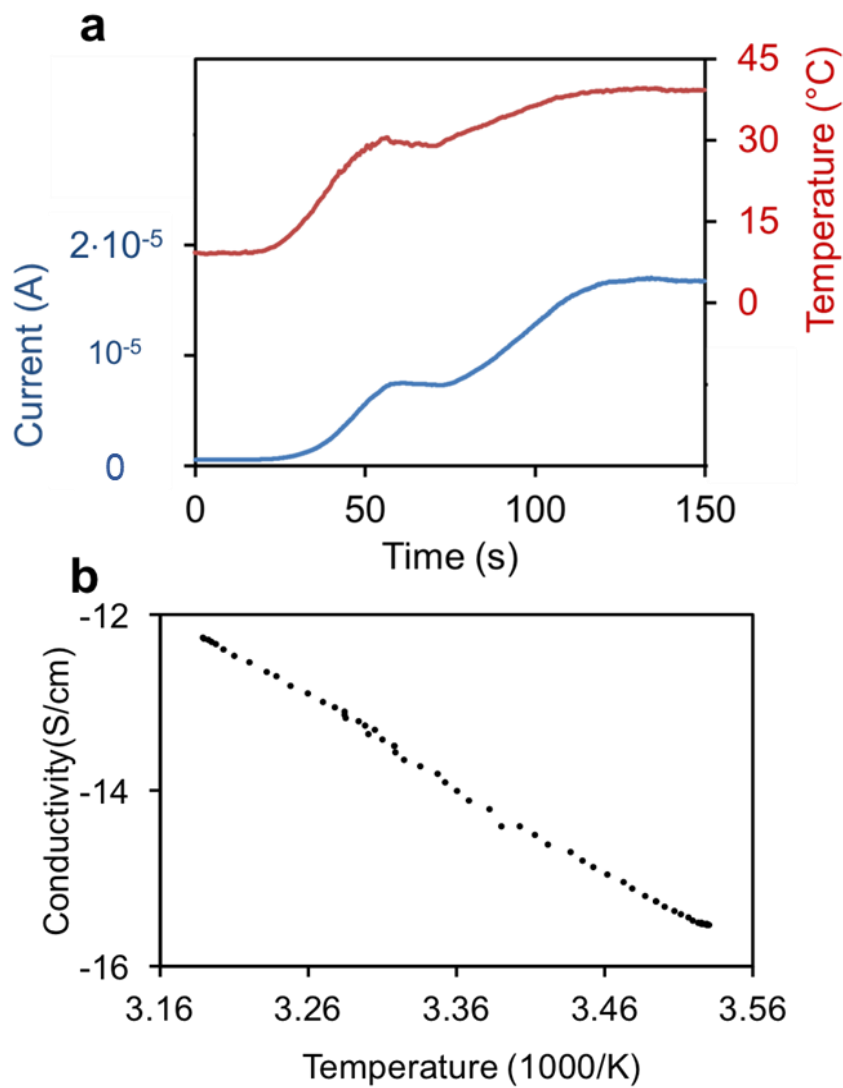

**Figure S3│Current-temperature characteristics of a pectin film and corresponding Arrhenius plot. a,** Temperature and current on a pectin film as a function of time. The blue line reports the current measured in the film. The red line represents the temperature of the film measured by the thermal camera. **b,** Arrhenius plot of electrical conductivity derived from Fig. S3a.

Figure S3b shows the Arrhenius plot of the conductivity of the pectin film. The activation energy can be derived from the slope of the line in the Arrhenius plot, multiplying it by the gas constant, as follows:

$$E_a = \text{Slope} \times R = 9.85 \text{ K} \times 8.31 \text{ J K}^{-1} \text{ mol}^{-1} = 81.9 \text{ kJ mol}^{-1}$$

## Large area sensors

Supplementary Video 1 shows an experiment with a large area sensor. A cup of coffee (covering less than a ninth of the film's surface area) was placed in the middle of the film while electrical current was monitored as a function of time, in parallel to the acquisition by a thermal camera and an optical video. To explain the current measurements, we associated a lumped element model to the sensor area. The resistance of the uniform film is concentrated into nine resistors (3 branches of 3 resistors in series, as shown in the white schematic diagram superimposed to the digital images in the video) connecting the two electrodes (cyan lines in the video). We assumed that only the resistor under the cup is contributing to the current increase, in proportion to the area under the cup and to its temperature variation (see thermal images in the video). We also assumed that the resistance change under the cup corresponds to the measurements reported in Fig. S3a, from which we derived that $R_{(30\ °C)} = 0.36 \times R_{(22\ °C)}$. We measured a current change across the large area of $\Delta I = 6.6\%$ between 22 and 30 °C. This value agrees with the assumption that the material has the same sensitivity as the one measured in the small films (Fig. 1A,E). The high responsivity of the pectin hydrogels is responsible for the ability to capture local events in large films, though without identifying the precise location and spatial dimension of the temperature source.

**Video S1│Large area film testing.** Results of an experiment performed on a sample A4 paper size with carbon tape electrodes, supported by a glass substrate and covered with an acetate sheet. In the movie, at time frame, $t = 1$ sec, we show on the left a flexible film sample, and on the right the same sample deposited on a glass slide for testing. At time $t = 14$ sec, we show imaginary lines dividing the sensor in 9 areas and the corresponding lumped elements model composed of 3 parallel branches of 3 resistors in series. At time $t = 19$ sec, we show that the resistor associated to the area heated by the cup changes color, to represent its dependence on temperature. At time $t = 30$ sec, we show in parallel a thermal camera video (on the top left corner), the current measured across the sensor (in the top right corner), and the corresponding optical video with superimposed schematic of the electrical model (bottom left corner). At time $t = 52$ sec we show the change in resistance estimated from Fig. S3a, which explains the change in current measured across the film, highlighted in cyan in the current-time plot.

# Bending

Figure S4 shows the bending positions corresponding to experiment in Fig. 3A.

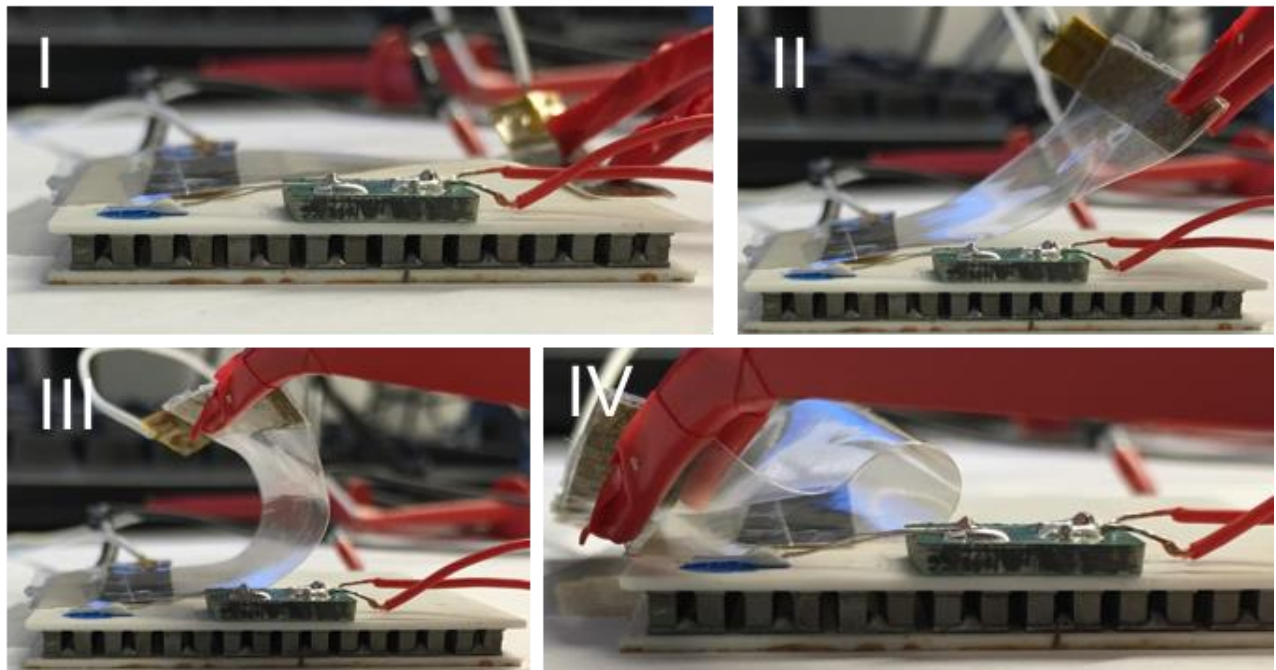

**Figure S4 Bending positions corresponding to experiment in Fig. 3A**

# Temperature mapping skins

Figure S5 shows the setup for the temperature measurement of films with multiple electrical contacts (herein referred to as "skins"). Figure S5a shows a 4-pixel configuration. The architecture of the skin is reported in Fig. 3C. The bottom insulator is made of silicon dioxide thermally grown on a silicon wafer, while the top insulator is a PDMS layer or acetate sheet. Chromium/gold contacts were sputtered on the bottom insulator. The same configuration was used for the 16-pixel skin, shown in Fig. S5b.

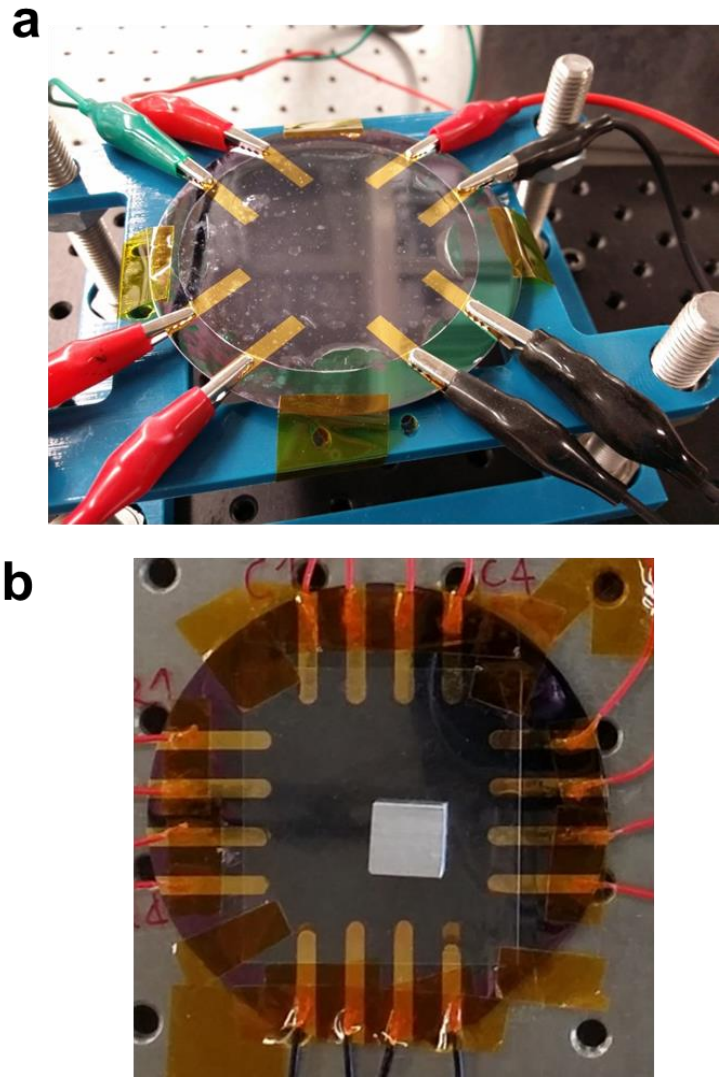

**Figure S5 Setup for the electrical measurements performed on skins. a,** 4-pixel skin. **b,** 16-pixel skin.

**Read out circuit**

Figure S6 shows the schematic of the read out circuit used for the 4-pixels skin. Figure S6a shows a diagram of the working principle: when temperature increases in correspondence of the pixel P11, the electrical resistance decreases locally. Thanks to the high sensitivity, the variation of resistance affects the series of resistances of the entire row and column as in Fig. S6a. This principle is similar to the one showed in Supplementary Video 1. Thus the local change in resistance is conveniently measurable interrogating each row and column in a separate time interval. This is done by allowing current to circulate in the row or column by switching on each CMOS couple. Only one row or column is on at any given time. Figure S6c shows an example of row 1 being read. Figure S6d shows the signals used to enable the reading. A buffer and an amplifier are used to condition and transfer the signals from the skin to the DAQ board. The DAQ read-out occurs when the enabling signal reaches its maximum value.

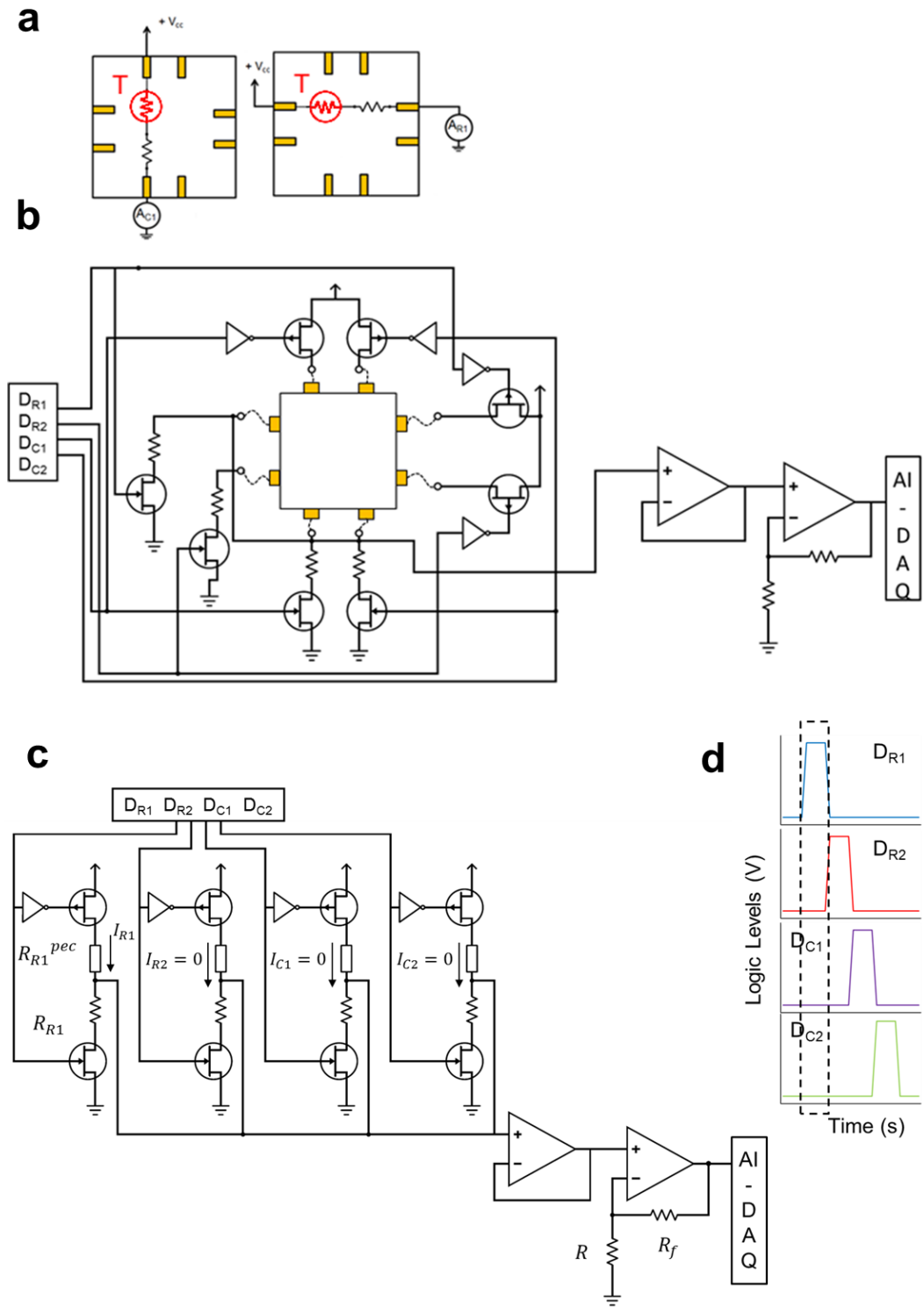

**Figure S6 Read out circuit for the 4-pixel skin. a,** Working principle. **b,** Skin connection. **c,** electrical schematic of the read out circuit when only one row is active. **d,** Enabling signals.

As responsivity and sensitivity are intrinsic characteristics of the material the only non-uniformity between rows columns is in their absolute current level. This is due to a non-perfectly uniform thickness of deposition of the material on its substrate. Prior to performing the experiments reported in Fig. 3D,E the absolute current levels of each row and column were leveled at room temperature, via the external resistors of the readout circuit in Fig. S6c.

**Measurement results on a 4-pixel temperature mapping skin**

We report in Tab. S1 and Fig. S7 the results obtained from measurements on a 4-pixel temperature mapping skin. To obtain these measurements and locally increase the temperature on each pixel, we position a finger for ~2 sec on a different quadrant of the skin. Table S1 shows the values representing the product of the signals in Fig. S7, when the finger is in each of the 4 positions ($P_{11}$, $P_{12}$, $P_{21}$, $P_{22}$). The table is obtained considering only the maximum voltage variation for each row and column ($R_1$, $R_2$, $C_1$, $C_2$). These values are multiplied as follows: $R_1C_1$, $R_1C_2$, $R_2C_1$, $R_2C_2$, the results are normalized for each experiment and reported in the schematics of Tab. S1. The sampling rate was 100 s$^{-1}$ and the averaging was 5 points per channel.

**Table S1 Values corresponding to the block diagrams with color map shown in Fig. 3D.**

| $P_{11}$ | C1 | C2 |
|---|---|---|
| R1 | 1 | 0.4 |
| R2 | 0.6 | 0.2 |

| $P_{12}$ | C1 | C2 |
|---|---|---|
| R1 | 0.4 | 1 |
| R2 | 0.2 | 0.5 |

| $P_{21}$ | C1 | C2 |
|---|---|---|
| R1 | 0.5 | 0.2 |
| R2 | 1 | 0.5 |

| $P_{22}$ | C1 | C2 |
|---|---|---|
| R1 | 0.3 | 0.5 |
| R2 | 0.5 | 1 |

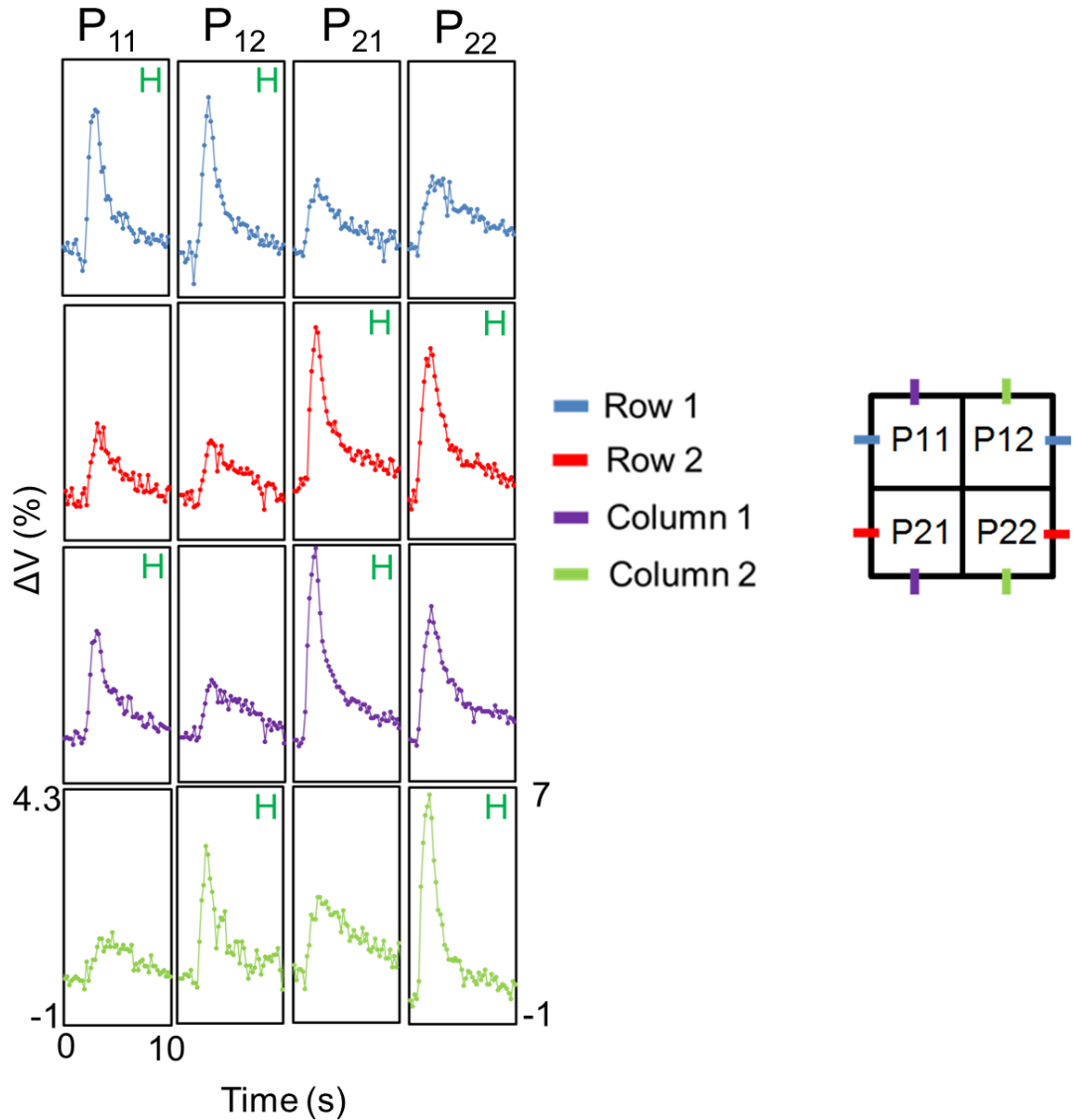

**Figure S7 Voltage at the readout circuit for every row and column in a 4-pixel skin**. Light blue: signals from row 1. Red: signals from row 2. Violet: signals from column 1. Green: signals from column 2. All the time scales in the panels are between 0 and 10 sec. All amplitudes are between -1 and 4.3 except for $P_{22}$, column 2, which is between -1 and 7 (as indicated in the corresponding panel). All signals for each finger position are acquired synchronously. The panels indicated by the letter "H" represent the highest signals for each row and column.

Figure S8 shows the current as a function of time in row 1 when $P_{11}$ is touched for ~1 sec. The noise level was undetectable, confirming that noise in the measurements of Fig. S7 was due to the readout circuit.

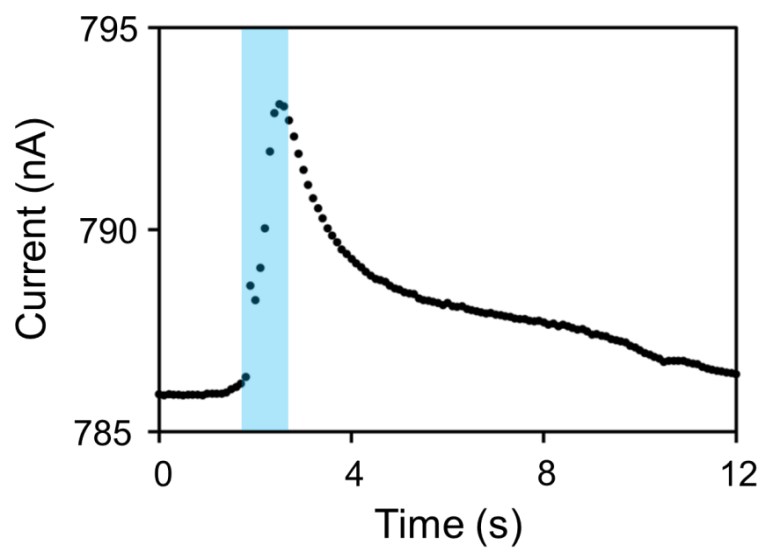

**Figure S8 Current vs. time when the film is touched with a finger.** The black dots report the current measurements. The blue shadowed region represents the finger's contact time on the skin.

Figure S9 shows the thermal fingerprint on the film, as acquired by the thermal camera. The image was taken immediately after lifting the finger that was in touch with the skin for ~1 sec. The temperature increase caused by the finger was less than 1 K, compared to temperature of the rest of the skin.

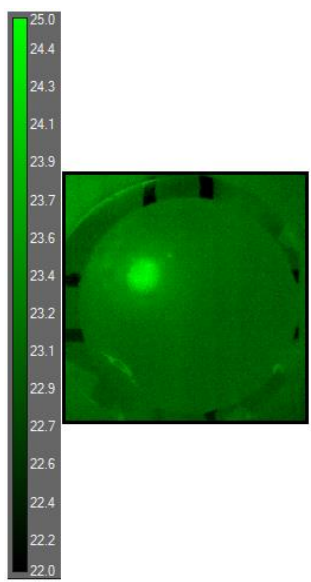

**Figure S9 Thermal image of the skin just after being touched with a finger.** The image shows the temperature map obtained with the thermal camera after a finger touched the skin in position $P_{11}$ (light green area). All values in the color intensity bar are expressed in °C.

To prove that the sensitivity of the skin is due to temperature changes and not pressure, we pressed the skin with a rounded metal part for ca. 5 sec, as shown in Fig. S10 (red panel). The skin was then touched with a finger, for the same length of time (green panel). When the skin was in contact with the metal tip, which was approximately at the same temperature of the skin, its effect on the current-time plot was negligible. However, when the skin was in contact with the finger, which was warmer than the skin, a current increase was evident.

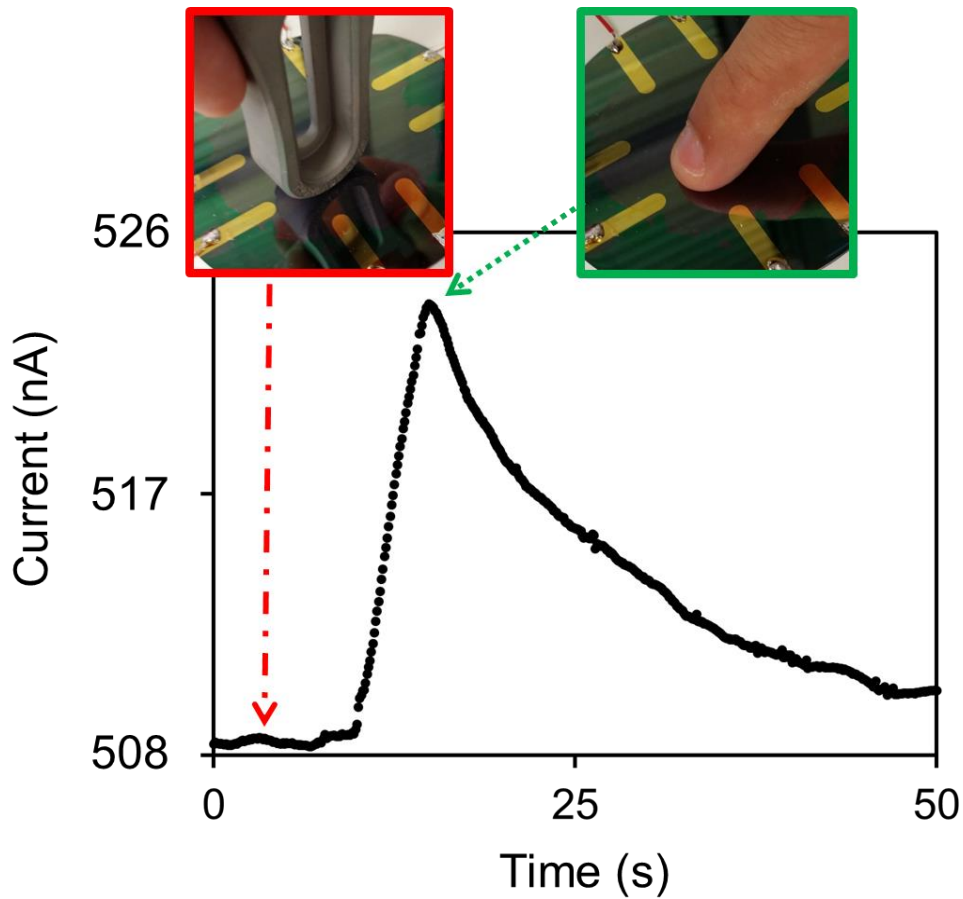

**Figure S10 Effect of pressure and temperature on a 4-pixel skin.**

**Measurement results on a 16-pixel temperature mapping skin**

We tested a 16-pixel skin, as shown in Fig. S5b, placing a warm (26 °C) aluminum square in contact with the skin. Figure S11 shows the thermal image of the skin, obtained after 0.8 sec. This image corresponds to the pixelated picture in Fig. 3F. Table S2 shows the values obtained from the skin, reported in Fig. 3E. To obtain these values, the acquisition rate used was 10 samples/sec, which corresponds to 1.25 samples/sec/channel.

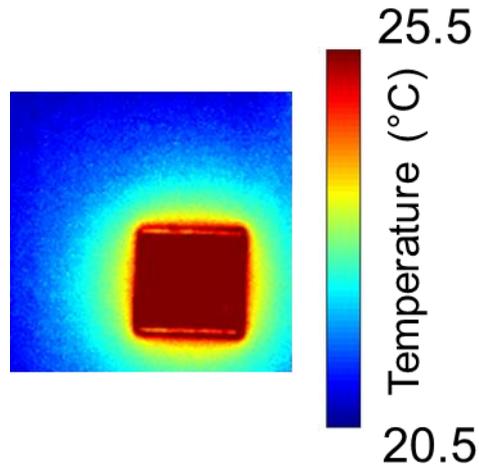

**Figure S11** Aluminum square in contact with the 16-pixel skin.

**Table S2** Values corresponding to the block diagram with color map shown in Fig. 3E.

|    | C1   | C2   | C3   | C4   |
|----|------|------|------|------|
| R1 | 0.45 | 0.49 | 0.59 | 0.59 |
| R2 | 0.59 | 0.63 | 0.76 | 0.76 |
| R3 | 0.77 | 0.82 | 1.00 | 0.99 |
| R4 | 0.74 | 0.79 | 0.96 | 0.95 |

### a.c. measurements

Figure S12 shows a typical thermal response measured at three different frequencies on a pectin film samples. The current is reported in arbitrary units since it was measured as the RMS value of the voltage drop on a resistor (50 kΩ) in series with the sample (see inset of Fig S12). No responsivity difference between measurements at different frequencies and d.c. measurements was found.

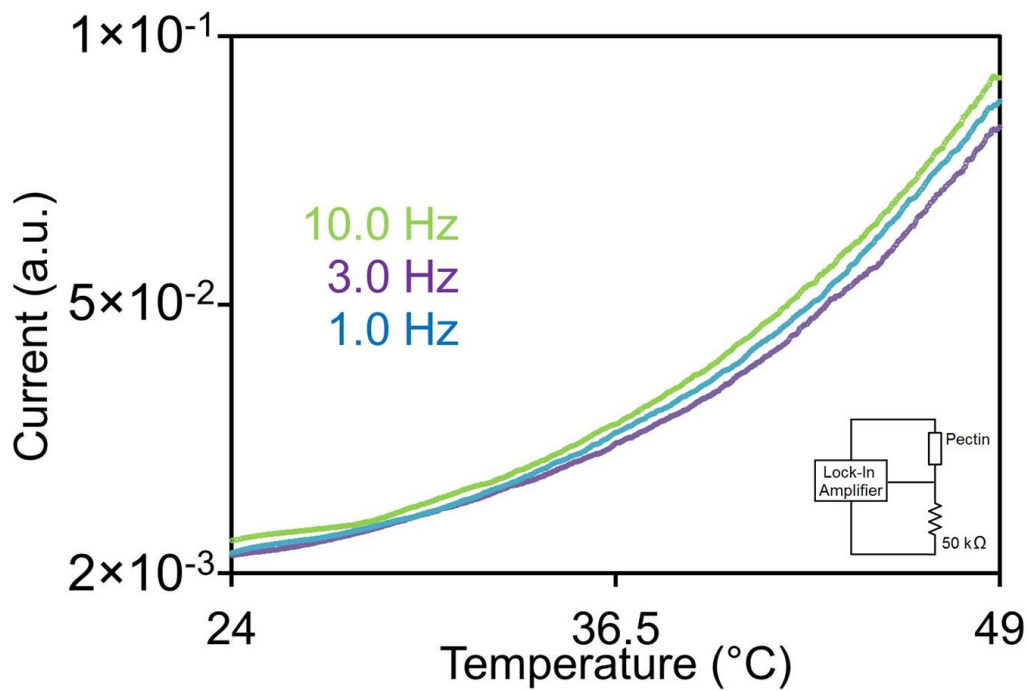

**Figure S12 Alternating current measurements on the pectin films.**